\documentclass[aps,preprint,showpacs,amsmath,amssymb,epsfig]{revtex4}
\usepackage{graphicx}
\usepackage{bm}

\begin{document}

\title{Higher Twists and Color Polarizabilities in the Neutron}

\author{
Z.-E.~ Meziani,$^{\ntemple}$
W.~Melnitchouk,$^{\njlab}$
J.-P.~Chen,$^{\njlab}$
Seonho~Choi,$^{\ntemple}$
T.~Averett,$^{\njlab,\nwm}$
G.~Cates,$^{\nuva}$
C.W.~de~Jager,$^{\njlab}$
A.~Deur,$^{\njlab,\nuva}$
H.~Gao,$^{\nmit}$
F.~Garibaldi,$^{\ninfn}$
R.~Gilman,$^{\njlab,\nrutgers}$
E.W.~Hughes,$^{\ncaltech}$
X.~Jiang,$^{\nrutgers}$
W.~Korsch,$^{\nkentucky}$
K.~Kramer,$^{\nwm}$
N.~Liyanage,$^{\nuva}$
K.~Slifer,$^{\ntemple}$
J.-C.~Yang$^{\nchungham}$
}
\affiliation{
\baselineskip 2 pt
\vskip 0.1 cm
\centerline{{$^{\ntemple}$Temple University, Philadelphia, Pennsylvania
19122}}	
\centerline{{$^{\njlab}$Thomas Jefferson National Accelerator Facility,
Newport News, Virginia 23606}}	
\centerline{{$^{\nwm}$The College of William and Mary, Williamsburg,
Virginia 23187}}		
\centerline{{$^{\nuva}$University of Virginia, Charlottesville, Virginia
22904}}
\centerline{{$^{\nmit}$Massachusetts Institute of Technology, Cambridge,
Massachusetts 02139}}
\centerline{{$^{\ninfn}$Istituto Nazionale di Fiscica Nucleare, Sezione
Sanit\`a, 00161 Roma, Italy}}
\centerline{{$^{\nrutgers}$Rutgers, The State University of New Jersey,
Piscataway, New Jersey 08855}}
\centerline{{$^{\ncaltech}$California Institute of Technolgy, Pasadena,
California 91125}}
\centerline{{$^{\nkentucky}$University of Kentucky, Lexington, Kentucky
40506}}
\centerline{{$^{\nchungham}$Chungnam National University, Taejon 305-764, Korea}}
}

\newcommand{\ntemple}{1}
\newcommand{\njlab}{2}
\newcommand{\nwm}{3}
\newcommand{\nmit}{5}
\newcommand{\nuva}{4}
\newcommand{\ninfn}{6}
\newcommand{\nrutgers}{7}
\newcommand{\ncaltech}{8}
\newcommand{\nkentucky}{9}
\newcommand{\nchungham}{10}


\begin{abstract}
Color polarizabilities of the neutron are extracted from data on the
lowest moment of the spin-dependent $g_1$ structure function.
New data in the resonance region from Jefferson Lab at
$Q^2 \alt 1$~GeV$^2$, in combination with world data at higher $Q^2$,
allow a systematic determination of the $1/Q^2$ corrections,
and provide the first constraints on $1/Q^4$ corrections.
The results suggest that higher-twist effects in the neutron are
small, and that quark-hadron duality may be approximately
valid, even down to $Q^2 \sim 1$~GeV$^2$.
\end{abstract}


\maketitle

The nature of quark confinement in QCD remains a central mystery of
strong interaction physics.
Any satisfactory solution to this problem requires understanding of
the dynamics of quark and gluon interactions at large distances, where
QCD becomes highly nonperturbative.

The properties of the quark and gluon constituents of the nucleon are
usually determined by examining how the quark and gluon fields respond
to external probes.
Deep inelastic electron scattering at high energy transfer $\nu$
and momentum transfer squared $Q^2$
has provided a wealth of information on the single-particle momentum
and spin distributions of quarks and gluons in the nucleon.
Interactions between quarks and gluons at short distances, which give
rise to scaling violations, can be described perturbatively through the
$Q^2$ evolution equations.
To probe directly long-range quark-gluon interactions, on the other
hand, requires the study of structure functions at intermediate values
of $Q^2$, in the pre-asymptotic region.

In this region a transition takes place between deep inelastic
scattering at high $Q^2$, where structure functions are determined by
incoherent scattering from partons, and the low $Q^2$ region which is
characterized by 
resonance dominance of structure functions~\cite{SVZ}.
The region of intermediate $Q^2$ ($Q^2 \sim 1$~GeV$^2$) exposes
quark-gluon correlation effects, which are otherwise suppressed
in the scaling regime, while still allowing an expansion in $1/Q^2$
to be meaningful.
An example of such effects is the so-called color polarizabilities
(not to be confused with the more familiar electromagnetic
polarizabilities), which describe how the color electric and magnetic
fields respond to the spin of the nucleon.

In this paper we analyze data from Jefferson Lab Hall~A experiment
E94-010 \cite{E94010} on the neutron spin dependent structure
function $g_1^n$ at $Q^2 \alt 1$~GeV$^2$, which when consistently
combined with the world high $Q^2$ data, enable a first accurate
determination of the color polarizabilities in the neutron.
In addition, we explore the closely related phenomenon of quark-hadron
duality, where the $g_1$ spin structure function in the resonance
region averages to the scaling function measured at high $Q^2$.
This duality has been explored in recent experiments for the proton and
deuteron structure functions \cite{DUALF2,HERMESA1}, but has not been
studied in the case of neutrons.

The analysis makes use of the operator product expansion (OPE) in QCD,
where at large $Q^2$ the lowest moment, $\Gamma_1^n(Q^2)$, of the spin
structure function $g_1$ of the neutron is expanded in inverse powers
of $Q^2$,
\begin{eqnarray}
\label{eq:twist}
\Gamma_1^n(Q^2)
&\equiv& \int_0^1 dx\ g_1^n(x,Q^2)\
=\ \sum_{\tau=2,4,\cdots} {\mu_\tau^n(Q^2) \over Q^{\tau-2}}\
\end{eqnarray}
with the coefficients $\mu_\tau^n$ related to nucleon matrix elements
of operators of twist $\leq \tau$.
Here twist is defined as the mass dimension minus the spin of an
operator, and $x = Q^2/2M\nu$ is the Bjorken $x$ variable, with
$M$ the neutron mass.
For each twist the $Q^2$ dependence in $\mu_\tau^n$ can be calculated
perturbatively as a series in $\alpha_s$.
Note that the application of the OPE requires summation over all
hadronic final states, including the elastic at $x=1$.

The leading-twist (twist-2) component, $\mu_2^n$, is determined by
matrix elements of the axial vector operator
$\bar\psi \gamma_\mu \gamma_5 \psi$, summed over various quark flavors.
It can be decomposed into flavor triplet ($g_A$), octet ($a_8$) and
singlet ($\Delta\Sigma$) axial charges,
\begin{eqnarray}
\label{eq:mu2}
\mu_2^n(Q^2)
&=& C_{\rm ns}(Q^2)
  \left( - {1 \over 12} g_A\ +\ {1 \over 36} a_8 \right)
+ C_{\rm s}(Q^2) {1 \over 9} \Delta\Sigma\ ,
\end{eqnarray}
where $C_{\rm ns}$ and $C_{\rm s}$ are the nonsinglet and singlet
Wilson coefficients \cite{LARIN}.
%
%
%
The nonsinglet triplet axial charge is obtained from neutron
$\beta$-decay, $g_A\ =\ 1.2670(35)$ \cite{PDG}, while the octet axial
charge is extracted from hyperon weak decay matrix elements assuming
SU(3) flavor symmetry, $a_8\ =\ 0.579(25)$ \cite{PDG}.
In order to factorize all of the $Q^2$ dependence into the Wilson
coefficients, we use the renormalization group invariant definition of
the matrix element of the singlet axial current,
$\Delta\Sigma \equiv \Delta\Sigma(Q^2=\infty)$.

The higher-twist contribution to $\Gamma_1^n(Q^2)$ can be obtained
by subtracting the leading-twist term from the total,
\begin{eqnarray}
\label{eq:DelGam}
\Delta\Gamma_1^n(Q^2)
&\equiv& \Gamma_1^n(Q^2) - \mu_2^n(Q^2)\
 =\ { \mu_4^n (Q^2) \over Q^2 }\
+\ { \mu_6^n (Q^2) \over Q^4 }\
+\ {\cal O}\left( {1\over Q^6} \right) .
\end{eqnarray}
The coefficient of the $1/Q^2$ term contains a twist-2 contribution,
$a_2^n$, and a twist-3 term, $d_2^n$, in addition to the genuine
twist-4 component, $f_2^n$ \cite{SV,MANK,JI_CHI},
\begin{eqnarray}
\label{eq:mu4}
\mu_4^n
&=& {1 \over 9} M^2
\left( a_2^n + 4 d_2^n + 4 f_2^n \right)\ .
\end{eqnarray}
The twist-2 term $a_2^n$ arises from the target mass correction
\cite{JU_G1}, and is related to the second moment of the twist-2
part of the $g_1^n$ structure function.
%
%
%
%

Interaction terms appear through the twist-3 and twist-4 contributions.
The coefficient $d_2^n$ is given by the twist-3 part of the
$x^2$-weighted moment of the $g_2^n$ structure function, and corresponds
to a matrix element of an operator which involves both quark and gluon
fields \cite{JU_G1}.

The twist-4 contribution to $\mu_4^n$ is defined by the
matrix element
\begin{eqnarray}
\label{eq:f2op}
f_2^n\ M^2 S^\mu
&=& {1 \over 2} \sum_i e_i^2\
\langle N |
 g\ \bar\psi_i\ \widetilde{G}^{\mu\nu} \gamma_\nu\ \psi_i
| N \rangle\ ,
\end{eqnarray}
where $\widetilde{G}^{\mu\nu}$ is the dual gluon field strength tensor,
and $g$ is the strong coupling constant.

The twist-3 and 4 operators describe the response of the collective
color electric and magnetic fields to the spin of the neutron.
Expressing these matrix elements in terms of the components of
$\widetilde{G}^{\mu\nu}$ in the nucleon rest frame, one can relate
$d_2^n$ and $f_2^n$ to color electric and magnetic polarizabilities.
These are defined as \cite{MANK,JI_CHI}
\begin{eqnarray}
\label{eq:chiE}
\chi_E^n\ 2 M^2 \vec S
&=& \langle N |\ \vec j_a \times \vec E_a\ | N \rangle\ ,  \\
\label{eq:chiB}
\chi_B^n\ 2 M^2 \vec S
&=& \langle N |\ j_a^0\ \vec B_a\ | N \rangle\ ,
\end{eqnarray}
where $\vec S$ is the neutron spin vector,
$j_a^\mu = -g \bar\psi\gamma^\mu t_a \psi$ is the quark current,
and $\vec E_a$ and $\vec B_a$ are the color electric and
magnetic fields, respectively.
In terms of $d_2^n$ and $f_2^n$ the color polarizabilities can be
expressed as
\begin{equation}
\chi_E^n = {2 \over 3} \left( 2 d_2^n\ +\ f_2^n \right)\ ,  \ \ \
\chi_B^n = {1 \over 3} \left( 4 d_2^n\ -\ f_2^n \right)\ .
\end{equation}
%


An analysis of the twist-4 matrix elements was carried out in
Ref.~\cite{JM} using data on $\Gamma_1^n$ from the E143 experiment
at SLAC \cite{E143} at $Q^2 = 0.5$ and 1.2~GeV$^2$.
Subsequently, SLAC experiments E154 \cite{E154} and E155 \cite{E155},
the HERMES Collaboration at DESY \cite{HERMES}, and the SMC at CERN
\cite{SMC} have collected data on $\Gamma_1^n$ over a larger range of
$Q^2$.
The precision of the earlier data in the resonance region was rather
poor, and no data existed for $0.5 < Q^2 < 1$~GeV$^2$, which precluded
the size of the twist-4 matrix elements from being well constrained.
More recently, precision data from Jefferson Lab Hall~A experiment
E94-010 \cite{E94010} were collected on $\Gamma_1^n$ at several $Q^2$
values below 1~GeV$^2$.
These data allow one for the first time to determine systematically
the size of the $f_2^n$ matrix element, and estimate the $1/Q^4$
correction.

Before proceeding with the extraction, it is important to ensure that
the various data sets are consistent with each other, in terms of the
assumptions and extrapolations beyond the measured region employed
in each analysis.
For instance, each of the experimental analyses uses a different
extrapolation to $x = 0$, which makes a naive comparison of the
published $\Gamma_1^n$ values problematic.
To this end we have reexamined all of the previous data on $\Gamma_1^n$
using the same set of assumptions for the $x \to 0$ as well as the
$x \to 1$ extrapolations.

To evaluate $\Gamma_1^n$ for each experiment, we added to the quoted
$\Gamma_1^n$ value from the measured region a contribution of the
unmeasured low-$x$ region using the Bianchi-Thomas (BT) parametrization
\cite{BT} up to $W^2 = 1000$~GeV$^2$.
This method was applied to the SLAC experimental results
\cite{E143,E154,E142}.
In the case of the HERMES and Jefferson Lab experiments, the
published $\Gamma_1^n$ values already include an estimate from the
unmeasured low-$x$ region using the BT parametrization, while for
the SMC experiment data exist above $W^2 = 1000$~GeV$^2$.
Except for the HERMES data, the remaining contribution down to $x=0$
was then evaluated assuming $g^n_1$ to be constant:
$g^n_1(x,Q^2) = g_1^{\rm BT}(x_{\rm min},Q^2)$ for the SLAC and JLab
experiments, with $x_{\rm min}$ defined for each $Q^2$ by
$W^2= 1000$~GeV$^2$, and
$g^n_1(x,Q^2) = g_1^{\rm SMC}(x_{\rm min}=0.003,Q^2=10~{\rm GeV}^2)$
for the SMC experiment.
Since the SLAC E155 Collaboration \cite{E155} only presents a moment
deduced from a next-to-leading order (NLO) analysis using world data,
we do not include this result in our fit.
Nevertheless, the NLO result for $\Gamma_1^n$ at $Q^2 = 5$~GeV$^2$
including the SLAC E155 data is very close to the result of the
SLAC E154 NLO analysis \cite{NLO_E154}.

To include the unmeasured $x \to 1$ region, we used a new fit of the
world neutron data which includes the high-precision $A_1^n$ data of
the Jefferson Lab E99-117 experiment \cite{A1N} at large $x$.
Since the elastic contribution is included separately, the maximum
value of $x$ is defined for each experiment by the pion
electroproduction threshold.
The resulting total moments $\Gamma_1^n$ from the world data are
plotted in Fig.~1 for $0.5 < Q^2 \leq 10$~GeV$^2$, where the total
uncertainty in each data set is the quadratic sum of the statistical and
systematic uncertainties.
The Jefferson Lab experiment E94-010 (filled circles) extends the range
of $Q^2$ with precision data below $Q^2 = 1$~GeV$^2$.
In all cases the data include both the inelastic and elastic
contributions, with the latter taken from the fit in Ref.~\cite{MMD}.

\begin{figure}[ht]
\label{fig:1}
\begin{center}
\hspace*{-0.5cm}
\centerline{\includegraphics[scale=0.8]{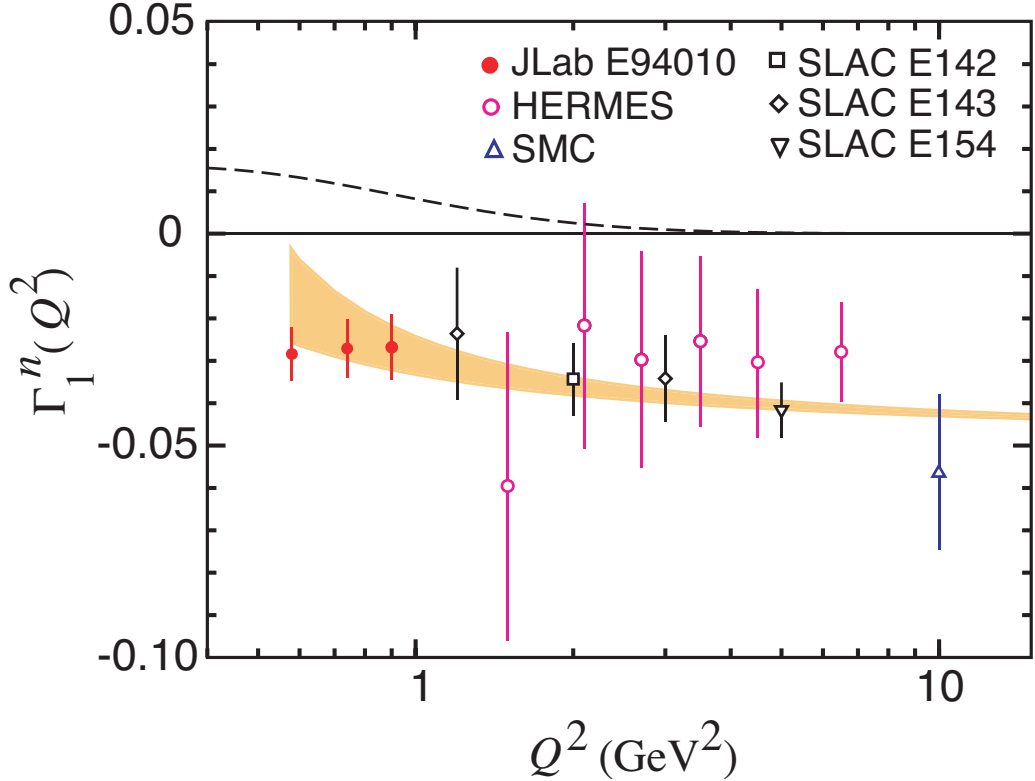}}
\end{center}
\vspace*{-0.5cm}
\caption{$Q^2$ dependence of $\Gamma_1^n$ from various experiments.
   The error bars are a quadratic sum of statistical and
   systematic uncertainties.
   The twist-2 contribution from Eq.~(\protect\ref{eq:mu2})
   is given by the band with $\Delta\Sigma=0.35$, and its width
   represents the uncertainty in $\alpha_s$.
   The elastic contribution is indicated by the dashed curve.}
\end{figure}

The twist-2 contribution $\mu_2^n$ is determined by fitting the neutron
data in Fig.~1 assuming there are no higher twists beyond
$Q^2 = 5$~GeV$^2$, from which we obtain $\Delta\Sigma=0.35\pm 0.08$,
where the uncertainty is statistical.
Using this central value, the twist-2 contribution is illustrated in
Fig.~1 by the shaded band, with the extrema representing the range of
uncertainty associated with the value of $\alpha_s$ in the Wilson
coefficients.
The exact value of $\Delta\Sigma$ depends somewhat on the $x \to 0$
behavior assumed in the extrapolation beyond the measured region.
However, since the higher-twist contributions are determined from
the relative variation in $\Gamma_1^n$ from high to low $Q^2$,
the absolute normalization of the leading-twist contribution does not
play a major role in determining $f_2^n$.

The higher-twist contribution $\Delta\Gamma_1^n$, obtained by
subtracting the leading-twist curves in Fig.~1 from data on the total
moment $\Gamma_1^n$, is shown in Fig.~2 as a function of $1/Q^2$ for
$\Delta\Sigma=0.35$.
Here we have used $a_2^n = -0.0031(20)$ for the target mass corrections,
obtained from a fit to the world neutron data \cite{A1N} at
$Q^2 = 5$~GeV$^2$, and the value $d_2^n = 0.0079(48)$ for the twist-3
matrix element obtained from SLAC experiment E155X~\cite{E155X}. At
this $Q^2$ value $a_2^n$ and $d_2^n$ are dominated by their leading-twist
contributions.

While the $Q^2$ evolution of the (twist-2) $a_2^n$ is straightforward,
the evolution of higher-twist structure functions is in general rather
more involved.
For the twist-4 $f_2^n$ matrix element the $Q^2$ evolution was computed
in Refs.~\cite{SV,KUYK} to leading logarithmic order.
In this analysis we assume the leading-twist values for $a_2^n$ and
$d_2^n$ at $Q^2=5$~GeV$^2$ and use the results from Refs.~\cite{SV,KUYK}
to account for the logarithmic $Q^2$ dependence of $f_2^n$.
In practice, the inclusion of $\alpha_s$ dependence of the $1/Q^2$
corrections has very little influence on the values of the higher
twists that we extract.

\begin{figure}[ht]
\label{fig:2}
\begin{center}
\hspace*{-0.5cm}
\centerline{\includegraphics[scale=0.8]{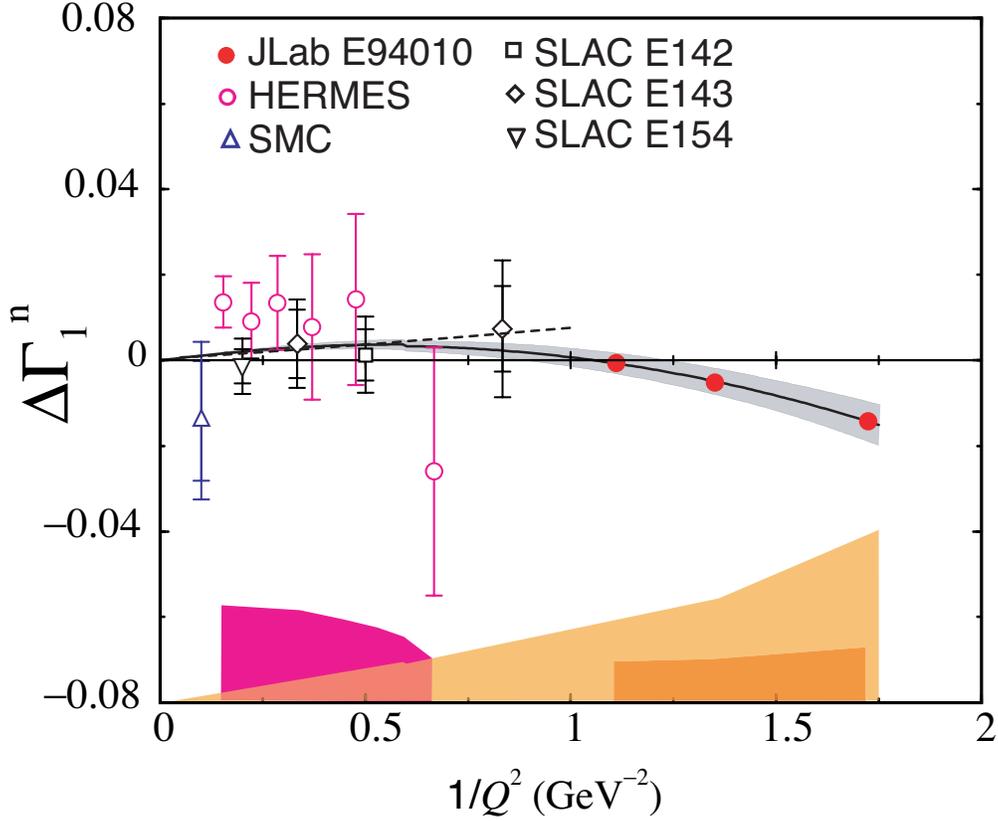}}
\end{center}
\vspace*{-0.5cm}
\caption{Higher-twist correction $\Delta\Gamma_1^n$ versus $1/Q^2$.
   The world data points include statistical (inner ticks) and total
   uncertainties (outer ticks), except for those of HERMES and JLab E94010,
   for which only statistical uncertainties are shown with error bars,
   with systematic uncertainties indicated by the dark bands at the
   bottom of the figure.
   The solid curve is a 2-parameter ($f_2^n$ and $\mu_6^n$) fit to the
   $Q^2 > 0.5$~GeV$^2$ data, while the dashed curve is a 1-parameter
   ($f_2^n$ only) fit to the $Q^2 > 1$~GeV$^2$ data.
   The band around the solid curve represents the uncertainty of the fit due
   to statistical uncertainties, and the light band at the bottom of
   the figure corresponds to the total uncertainty.}
\end{figure}

The solid curve in Fig.~2 represents a 2-parameter minimum $\chi^2$
fit to the $\Delta\Gamma_1^n$ data for $Q^2 > 0.5$~GeV$^2$, using
Eq.~(\ref{eq:DelGam}) with $f_2^n$ and the $1/Q^4$ correction $\mu_6^n$
as free parameters.
We neglect any possible $Q^2$ dependence in $\mu_6$ itself, which
should be a reasonable assumption within the present uncertainties.
The best fit values for the twist-4 and $1/Q^4$ corrections, using
only the statistical uncertainty for each experiment, are found to be
\begin{equation}
f_2^n = 0.033 \pm 0.005\ , \ \ \ \mu_6^n = (-0.019 \pm 0.002) M^4\ ,
\end{equation}
normalized at $Q^2=1$~GeV$^2$.
Including the total systematic uncertainty for each experiment, we find
\begin{equation}
f_2^n = 0.034 \pm 0.043\ , \ \ \ \mu_6^n = (-0.019 \pm 0.017) M^4\ .
\end{equation}
With this latter value of $f_2^n$, the overall $1/Q^2$ correction to
the leading-twist term, at a scale $Q^2=1$~GeV$^2$, including the
target mass, twist-3 and twist-4 contributions extracted from data,
is $\mu_4^n = (0.019 \pm 0.024) M^2$.
Combining this with the $1/Q^4$ term, the total higher-twist
contribution $\Delta\Gamma_1^n$ is almost exactly zero at
$Q^2 = 1$~GeV$^2$.

The reliability of truncating the expansion at $1/Q^4$ order down
to $Q^2 = 0.5$~GeV$^2$ can be tested by adding a $\mu^n_8/Q^6$ term.
Performing a 3-parameter fit results in almost identical values for
$f_2^n$ and $\mu_6^n$, with $\mu_8^n = (0.00 \pm 0.03) M^6$.
On the other hand, if the data are fitted with just the $1/Q^2$ term,
one generally finds a smaller value for $f_2^n$, namely
$f_2^n = -0.014 \pm 0.010$, with a somewhat larger $\chi^2$
(see also Ref.~\cite{KAO}).
This clearly illustrates the influence of the JLab E94-010 data,
which suggest a negative trend in $\Delta\Gamma_1^n$ for
$Q^2 < 1$~GeV$^2$.
Excluding the JLab results and fitting only the $Q^2 > 1$~GeV$^2$ data,
the uncertainty on $f_2^n$ becomes 2--3 times larger, and the $\mu_6^n$
coefficient essentially unconstrained.
A 1-parameter fit to the $Q^2 > 1$~GeV$^2$ data, indicated by the
dashed curve in Fig.~2, gives a slightly smaller $f_2^n$ value,
$f_2^n = 0.012 \pm 0.029$.

The extracted value of $f_2^n$ can be compared with the earlier analysis
in Ref.~\cite{JM}, which found $f_2^n = 0.07 \pm 0.08$ using a
1-parameter fit to $Q^2 > 1$~GeV$^2$ data.
The present analysis improves the uncertainty on $f_2^n$ by a factor
$\sim 3$ using the non-JLab data, and by a factor $\sim 8$ including the
JLab data.
Of course, the uncertainty on $f_2^n$ in Ref.~\cite{JM} would be
significantly larger had a 2-parameter fit been employed.

The value of $f_2^n$ extracted from the data can be compared with
nonperturbative models.
For example, QCD sum rules \cite{MANK,BBK} yield negative values
for the twist-4 matrix elements:

$\left. f_2^n\right|_{\rm sum\ rule} = -0.018(17)$ and $-0.013(6)$
from Refs.~\cite{BBK} and \cite{MANK}, respectively.
On the other hand, calculations within the instanton model give
\cite{INSTANTON}
$\left. f_2^n\right|_{\rm instanton} = 0.038$,
in good agreement with our central value.
Estimates of the $\tau=4$ matrix elements using the MIT bag model
\cite{JU_G1}, evolved from the bag scale, give
$\left. f_2^n\right|_{\rm bag} = 0.0$,
which is also consistent with our result.

Combining the extracted $f_2^n$ values with the $d_2^n$ value obtained
by the E155X Collaboration \cite{E155X}, the color electric and
magnetic polarizabilities, normalized at $Q^2=1$~GeV$^2$, are found
to be
\begin{equation}
\chi_E^n = 0.033 \pm 0.029\ , \ \ \ \chi_B^n = -0.001 \pm 0.016\ .
\end{equation}
These results indicate that both the color electric and magnetic
polarizabilities in the neutron are relatively small, with the magnetic
polarizability consistent with zero.
The small values of the higher-twist corrections suggest that the
long-range, nonperturbative interactions between quarks and gluons in
the neutron are not as dominant at values of $Q^2 \agt 0.5$~GeV$^2$
as one may have expected.
This would imply strong cancellations between neutron resonances
resulting in the dominance of the leading-twist contribution to
$\Gamma_1^n$.
The results presented here therefore provide a spectacular confirmation
of quark-hadron duality in spin-dependent structure functions.
A dedicated experiment \cite{NIL:01} to study duality in neutron spin
structure functions in the resonance region is currently being
analyzed.


In summary, we have determined the first moment of the neutron spin
structure function $g_1$ from the world data using consistently the
same extrapolation method to $x=0$ and $x=1$.
From these data we extracted the size of the twist-4 matrix element
$f_2^n$ of the neutron.
The neutron color polarizabilities were determined by combining $f_2^n$
and the twist-3 matrix element $d_2^n$ from SLAC experiment E155X,
which leads to a small and slightly positive value of $\chi_E^n$ and
a value of $\chi_B^n$ close to zero.
Future precision measurements~\cite{NIL:01,AVE:97,MEZ:03,JLA:00} of the
$g_1^n$ and $g_2^n$ structure functions at $Q^2 \approx 1-4$~GeV$^2$
will reduce the uncertainty in the extracted higher-twist coefficients,
as will better knowledge of $\alpha_s$ in the intermediate $Q^2$ region.

This work was supported by the U.S. Department of Energy (DOE) and the 
National  Science Foundation.  The Southeastern Universities
Research Association operates the Thomas Jefferson National Accelerator
Facility for the DOE under  contract DE-AC05-84ER40150.


\end{document}